\documentclass[aps,prl,preprint,groupedaddress]{revtex4-1}

\usepackage{amsmath}
\usepackage{amssymb}
\usepackage{epsfig}
\usepackage{color}
\usepackage{graphicx}

\def\be{\begin{equation}}
\def\ee{\end{equation}}
\def\beq{\begin{eqnarray}}
\def\eeq{\end{eqnarray}}
\def\n{\nonumber}

\usepackage{color}
\usepackage{graphicx}

\begin{document}


\title{Spherically symmetric isothermal fluids in $f(R,T)$ gravity}


\author{Sudan Hansraj}
\altaffiliation{} \affiliation{Astrophysics and Cosmology Research
Unit, University of KwaZulu Natal} \email[]{hansrajs@ukzn.ac.za}


\date{\today}

\begin{abstract}

We analyze the isothermal property in static fluid spheres within
the framework of the modified $f(R, T)$ theory of gravitation. The
equation of pressure isotropy of the standard Einstein theory is
preserved however, the energy density and pressure are expressed in
terms of both gravitational potentials. Invoking the isothermal
prescription requires that the isotropy condition assumes the role
of a consistency condition and an exact model generalizing that of
general relativity is found. Moreover it is found that the Einstein
model is unstable and acausal while the $f(R, T)$ counterpart is
well behaved on account of the freedom available through an
additional coupling constant. The case of a constant spatial
gravitational potential is considered and the complete model is
determined. This model is markedly different from its Einstein
counterpart which is known to be isothermal. Dropping the
restriction on the density and imposing a linear barotropic equation
of state generates an exact solution and consequently a stellar
distribution as the vanishing of the pressure is possible and a
boundary hypersurface exists. Finally we comment on the case of
relaxing the equation of state but demanding an inverse square
fall-off of the density - this case proves intractable.

\end{abstract}


\maketitle

\section{Introduction}

Phenomenological theories of gravity have been on the increase in
recent times. Such ideas purport to resolve the problems which are
shortcomings of the standard Einstein's general theory of
relativity. Specifically, the late time accelerated expansion of the
universe is not a natural consequence of general relativity with the
cosmological evolution equations suggesting a decelerating universe.
This is inconsistent with experimental data conveyed by  the
supernovae Type 1a data \cite{Perlmutter}, Baryon Acoustic
Oscillations \cite{Eisenstein} and the WMAP survey involving the
cosmic microwave background \cite{Spergel}. In order to resolve the
difficulty, proposals of exotic matter fields have emerged. These
include dark energy, dark matter, quintessence, phantom fields and
the like. These latter fields do not as yet enjoy any experimental
support even though their motivations may be  sound.

An alternative approach is to reconsider the fundamental geometry
prescripts. A modification of the action principle may have the
potential to resolve the anomalies with the standard theory. For
example, in $f(R)$ theories \cite{staro} the action involves a
polynomial in the Ricci scalar. It has been demonstrated that such
an approach may indeed explain the accelerated expansion of the
universe. It has been shown by Goswami {\it{et al}} \cite{goswami}
that the Buchdahl upperbound \cite{buch, Buchdahl1} for the
mass-radius ratio of general relativity may be improved in $f(R)$
theory with more matter per unit mass being admitted. The results
also have implications for our understanding of the dark matter
problem.  The serious drawback of $f(R)$ theory is the appearance of
higher derivative terms which correspond to ghosts. It is usual in
gravity theory to have at most second order equations of motion.
Moreover it has been demonstrated \cite{felice} that $f(R)$ theory
is conformally related to the  scalar-tensor field theory of Brans
and Dicke.

The most general tensor theory of gravity admitting at most second
order derivatives is the Lovelock theory \cite{lovelock1,lovelock2}.
The action consists of polynomials in the scalar invariants
constructed from  the Riemann tensor, Ricci tensor and the  Ricci
scalar. The drawback in this formalism is that the higher curvature
terms are only active in dimensions higher than 4. That is Lovelock
theory reduces to standard general relativity in dimensions 3 and 4
and makes a contribution to the dynamics from dimension 5 upwards. A
special case of the Lovelock polynomial is the second order term
known as the Gauss--Bonnet term that appears in the effective action
of heterotic string theory \cite{gross}. The exterior field for a
spherically symmetric star has been established by Boulware and
Deser \cite{boul} for the neutral sphere and by Wiltshire
\cite{wiltshire} for the charged case in the mid 1980s. However,
only recently were interior metrics found for perfect fluid
astrophysical objects \cite{hans-maha, maha-hans, chil-hans} in
Einstein--Gauss--Bonnet gravity that could be matched to the
Boulware-Deser \cite{boul} exterior metric.

If a scalar tensor action is sought then the most general such
theory yielding second order equations of motion is due to Horndeski
\cite{horn} and consists of the so called Fab Four components of the
effective lagrangian. Several studies into its cosmological
implications have been undertaken \cite{charmousis} and of late
compact objects such as black holes and neutron stars were
investigated by Silva {\it { et al}} \, \cite{silva}. Tensor
multi-scalar theory of gravity has also recently come into vogue
\cite{darmour}.

Harko  {\it {et al}} \cite{harko} have proposed an action that is a
function of the Ricci scalar $R$ and the trace of the energy
momentum tensor $T$ which goes by the name $f(R, T)$ gravity. The
equations of motion are indeed second order however the conservation
of energy is sacrificed. This is ostensibly a drawback of the
theory. However, it was argued by Rastall \cite{Rastall} that
spacetime curvature could account for non-compliance with the
Newtonian view of energy conservation
\cite{Rastall,Rastall1,Heydarzade}. Extensive investigations into
the $f(R, T)$ paradigm have been conducted  in recent times
\cite{correa}.

We examine the physically important case of perfect fluids
displaying the isothermal property that is an inverse square law
fall-off of density as well as a linear equation of state. In such
universes galaxies are considered as pointlike structures.  By
design such models can only describe cosmological fluids as no
hypersurface of vanishing pressure indicating a boundary is present.

The paper is structured as follows: Firstly we review the essential
ingredients of the $f(R, T)$ framework.  We then derive the
isothermal model in $f(R, T)$ theory and compare with the  solution
for Einstein gravity. In the next section we probe the consequences
of a constant gravitational potential since it is known in Einstein
gravity  that a necessary and sufficient condition for isothermal
behavior is a constant spatial gravitational potential.  Finally we
impose a linear barotropic equation of state on our model but
without any restriction on the density profile. Before we conclude
with a discussion, we comment on the case of an inverse square
fall-off of the density but without imposing an equation of state.

\vspace{0,5cm}
\section{Elements of $f(R,T)$ Theory}

The $f(R,T)$ gravity  action is given by
\begin{equation}
S=\frac{1}{16\pi}\int d^{4}xf(R,T)\sqrt{-g}+\int
d^{4}x\mathcal{L}_m\sqrt{-g}, \label{eq1}
\end{equation}
where $f(R,T)$ is an arbitrary function of the Ricci scalar $R$, and
$T$ is the trace of the energy momentum tensor $T_{\mu\nu}$. The
 Lagrangian density $\mathcal{L}_m$ for the matter field
is defined as
\begin{equation}
T_{\mu\nu}=
-\frac{2}{\sqrt{-g}}\frac{\delta\left(\sqrt{-g}\mathcal{L}_m\right)}{\delta
g^{\mu\nu}},\label{eq2}
\end{equation}
and its trace by $T=g^{\mu\nu}T_{\mu\nu}$.  The  Lagrangian density $\mathcal{L}_m$ of matter has the form
\begin{equation}\label{eq3}
T_{\mu\nu}=
g_{\mu\nu}\mathcal{L}_m-2\frac{\partial\left(\mathcal{L}_m\right)}{\partial
g^{\mu\nu}},
\end{equation}
and is dependent  only on the metric tensor components. Variation of
the action (\ref{eq1}) with respect to the metric $g^{\mu\nu}$
generates the field equations
\begin{eqnarray}\label{eq4}
f_R (R,T) R_{\mu\nu} - \frac{1}{2} f(R,T) g_{\mu\nu} + (g_{\mu\nu}\Box - \nabla_{\mu} \nabla_{\nu}) f_R (R,T)\nonumber \\
= 8\pi T_{\mu\nu} - f_{T(R,T)} T_{\mu\nu} -
f_{T(R,T)}\Theta_{\mu\nu},
\end{eqnarray}
where $f_R (R,T)= \partial f(R,T)/\partial R$ and $f_T
(R,T)=\partial f(R,T)/\partial T$. $\nabla_{\mu}$ denotes covariant
differentiation and the box operator $\Box$, is defined via
\begin{equation*}
\Box \equiv \partial_{\mu}(\sqrt{-g} g^{\mu\nu}
\partial_{\nu})/\sqrt{-g}, ~~~\text{and}~~~ \Theta_{\mu\nu}=
g^{\alpha\beta}\delta T_{\alpha\beta}/\delta g^{\mu\nu} .
\end{equation*}
The covariant divergence of Eq. (\ref{eq4})
 produces the equation
\begin{widetext}
\begin{eqnarray}\label{eq5}
\nabla^{\mu}T_{\mu\nu}&=&\frac{f_T(R,T)}{8\pi
-f_T(R,T)}[(T_{\mu\nu}+\Theta_{\mu\nu})\nabla^{\mu}\ln
f_T(R,T)\nabla^{\mu}\Theta_{\mu\nu}-(1/2)g_{\mu\nu}\nabla^{\mu}T].
\end{eqnarray}
\end{widetext}
which clearly shows that energy is not conserved in this system.
  With the help of  Eq. (\ref{eq3})  the quantity
$\Theta_{\mu\nu}$  is expressible as
\begin{equation}\label{eq6}
\Theta_{\mu\nu}=
-2T_{\mu\nu}+g_{\mu\nu}\mathcal{L}_m-2g^{\alpha\beta}
\frac{\partial^2 \mathcal{L}_m }{\partial g^{\mu\nu}\partial
g^{\alpha\beta}} .
\end{equation}
For the purposes of this investigation we consider a perfect fluid source with energy--momentum tensor
\begin{equation}\label{eq7}
T_{\mu\nu}=(\rho+p)u_\mu u_\nu-pg_{\mu\nu},
\end{equation}
where $p$ is the pressure and $\rho$ the energy density of strange
matter, with $u^{\mu}u_{\mu} = 1$ and $u^\mu\nabla_\nu u_\mu=0$. If we take the  matter Lagrangian density to be
$\mathcal{L}_m = -p$, and the Eq. (\ref{eq6}) we obtain the relationship
\begin{equation}\label{eq8}
\Theta_{\mu\nu}= -2T_{\mu\nu} -\rho g_{\mu\nu} .
\end{equation}
Following Harko {\it {et al}} we consider the simplest version of $f(R, T)$  namely $f(R,T)=R+2\chi T$ where $\chi$ is a coupling constant constant. The field equations are now given by
\begin{equation}\label{eq9}
G_{\mu\nu}=8\pi T_{\mu\nu}+\chi
Tg_{\mu\nu}+2\chi(T_{\mu\nu}+pg_{\mu\nu}),
\end{equation}
where $\chi$ can be positive or negative.   Eq. (\ref{eq5})  can now be written as
\begin{equation}\label{eq10}
(8\pi+2\chi)\nabla^{\mu}T_{\mu\nu}=-2\chi\left[\nabla^{\mu}(pg_{\mu\nu})+\frac{1}{2}g_{\mu\nu}\nabla^{\mu}T\right].
\end{equation}
and in the case of vanishing $\chi$ the law of energy conservation in Einstein gravity is recovered.

\section{ Field Equations}

In coordinates $(t, r, \theta, \phi)$ the most general spherically symmetric line element reads as
\begin{equation}
ds^2 = e^{\nu(r)}dt^2  - e^{\lambda (r)} dr^2 - r^2 \left(d\theta^2
+\sin^2 \theta d\phi^2\right), \label{eq11}
\end{equation}
 where $\nu(r)$ and $\lambda(r)$ are arbitrary functions of the radial
coordinate $r$ only. We consider a comoving fluid 4-velocity field $
u^a = e^{-\nu /2} \delta_0^a $ and
 a perfect fluid source with energy momentum tensor given in Eq. (\ref{eq4}). Additionally we use geometrized units
such that the gravitational constant $G$ and the speed of light $c$  are taken as unity.  Now  Eq. (\ref{eq9}) and Eq.
(\ref{eq11}) generate the field equations
\begin{eqnarray}
e^{-\lambda}\left(\frac{\lambda^\prime}{r}-\frac{1}{r^2}\right)+\frac{1}{r^2}&=&\left(8\pi+3\chi\right)\rho
-\chi p, \label{12}\\
e^{-\lambda}\left(\frac{\nu^\prime}{r}+\frac{1}{r^2}\right)-\frac{1}{r^2}&=&\left(8\pi+3\chi\right)p
-\chi \rho, \label{13}
\end{eqnarray}
where the prime denotes the derivative with respect to the radial
coordinate, $r$. Introducing the transformation  $e^{-\lambda} =
1-2m(r)/r$ we obtain
\begin{eqnarray}
m^{\prime}=4\pi r^2\rho+\frac{\chi(3\rho-p)r^2}{2},\label{eq14}
\end{eqnarray}
where the function $m = m(r)$ represents the gravitational mass. An additional equation may be written from
(\ref{eq10})
\begin{eqnarray}
\frac{dp}{dr}+(\rho+p)\frac{\nu^{\prime}}{2}=
\frac{\chi}{8\pi+2\chi}\left(p'-\rho^{\prime}\right),
\end{eqnarray}
that reduces to the energy conservation of general relativity when
$\chi = 0$. It is possible to rewrite  Eqs. (\ref{12}) and (\ref{13})
in terms of energy density ($\rho$) and pressure  ($p$)
in the form
\begin{widetext}
\begin{eqnarray}
 \rho &=& \frac{e^{-\lambda}}{((8\pi + 3\chi)^2-\chi^2)r^2} \left(2(\chi +
4\pi)(e^{\lambda}-1) +r((8\pi + 3\chi)\lambda' + \chi\nu')\right),
\label{5a} \\
p &=& \frac{e^{-\lambda}}{((8\pi + 3\chi)^2-\chi^2)r^2} \left(2(\chi
+ 4\pi)(1 - e^{\lambda}) + r(\chi\lambda' + (8\pi +
3\chi)\nu')\right) ,\label{5b}
\end{eqnarray}
\end{widetext}
while the equation of pressure isotropy $G^r_r = G^{\theta}_{\theta}$ reduces to
\begin{equation}
r^2(2\nu'' + \nu'^2 -\nu'\lambda') -2r(\nu' + \lambda')
+4(e^{\lambda} -1) = 0. \label{5c}
\end{equation}
Observe that the equation of isotropy is the same for the ordinary Einstein's
equations with a perfect fluid source. Therefore any  of the well
known solutions reported over the past century (for example see Delgaty and Lake \cite{Delgaty}) will satisfy (\ref{5c}).

\section{Solution of the field equations with the isothermal property }

A perfect fluid is said to be isothermal if the density and pressure both obey the inverse square law fall-off and consequently display the equation of state $p = \gamma \rho$ for some real number $0 < \gamma < 1$ \cite{saslaw}.
Accordingly let us insert
\be
p = \frac{A}{r^2} \hspace{0.5cm} {\mbox{and}} \hspace{0.5cm} \rho = \frac{B}{r^2}  \label{6}
\ee
where $A$ and $B$ are arbitrary parameters (at this stage) into equations (\ref{5a}) and (\ref{5b}). Observe that the field equations are essentially 3 in number and they contain four unknown functions. Accordingly, specifying two of the quantities, namely the density and pressure, appears to be over-determining the system. This is true, however, we shall utilise the pressure isotropy equation as a consistency condition and determine the relationship between $A$ and $B$ for the isothermal property to hold.
This is a similar route followed by Saslaw {\it{et al}} \cite{saslaw} in dealing with isothermal spheres in standard Einstein gravity.

Introducing (\ref{6}) into (\ref{12}) yields the differential
equation \be e^{-\lambda} (-1 + e^{\lambda} + r\lambda') = 8\pi B +
\chi (3B - A) \label{61}
 \ee which is written only in terms of the potential
$\lambda$. With the help of the substitution $e^{\lambda} = \beta
(r)$, equation (\ref{61}) assumes the form \be r\beta' + (1-w_1
)\beta^2 - \beta = 0 \label{62} \ee where we have set $w_1 = 8\pi B
+ \chi (3B-A)$. Equation (\ref{62}) is a Ricatti equation and is
solvable in the form \be \beta = e^{\lambda} = \frac{C_1 r}{C_1 r(
w_1 -1)-1} \label{63} \ee where $C_1$ is a constant of integration.
Putting (\ref{6}) into (\ref{13}) simplifies it to the form \be
e^{-\lambda} (1 - e^{\lambda} + r\nu') = w_2 \label{65} \ee where we
have labelled $w_2 = 8\pi A + \chi (3A - B)$.

Now inserting (\ref{63}) into (\ref{65}) generates the solution \be
e^{\nu} = \frac{\left(1-C_1 r (w_1 -1)\right){}^{\frac{w_2 +1}{1-
w_1}}}{r}  \ee

The isotropy equation (\ref{5c}) becomes \be  C_1 r \left(w_1^2+6
w_1 w_2+(w_2 -4) w_2\right)-w_1 -5 w_2 = 0 \ee and for consistency
it is required that the coefficient of $r$ and the constant term
simultaneously vanish. This is achieved for \be \{(w_1, w_2) \} =
\{(0; 0), (5, -1)\} \ee which translates to \be A = \frac{\chi - 4
\pi  }{4 \left(\chi ^2+6 \pi  \chi +8 \pi ^2\right)} \hspace{0.5cm}
{\mbox{and}} \hspace{0.5cm} B = \frac{7 \chi + 20 \pi }{4 \left(\chi
^2+6 \pi \chi +8 \pi ^2\right)} \ee expressing $A$ and $B$ in terms
of the coupling constant $\chi$.

To ensure a subluminal sound speed requires $0 < \gamma =
\frac{A}{B} < 1$ and this constrains the coupling constant to  \be
-4 \pi <\chi <-\frac{20 \pi }{7}  \ee for the stability of the
model. Moreover, this same interval guarantees that both density and
pressure remain positive for all radial values.   The mass of the
infinite sphere as \be m= \frac{\chi - 4 \pi  }{4 \left(\chi ^2+6
\pi  \chi +8 \pi ^2\right)} r + K  \ee for some constant $K$.
Observe that in the interval of validity above, the mass profile is
a monotonically increasing function.

\section{The Einstein isothermal model}

Setting $\chi = 0$ above regains the standard Einstein field
equations. Saslaw {\it {et al}} \cite{saslaw} obtained an isothermal
 cosmological model where the the geometric and dynamical variables
are expressed in terms of the parameter $\alpha$ which comes from
the equation of state $p = \alpha \rho$ and which corresponds to
$\frac{A}{B}$ in our formalism. We now make some observations about
this model. Setting $\chi = 0$ we get $A = -\frac{1}{8\pi}$ and $B =
\frac{5}{8\pi}$ thus giving the specific value $\alpha =
-\frac{1}{5}$. Importantly this value is negative showing that the
Saslaw model violates causality. In fact the constant value for the
metric potential $e^{\lambda} = 1 + \frac{4\alpha}{(1+\alpha)^2}$
evaluates to $-\frac{1}{4}$ which is not feasible. Accordingly the
Saslaw model is not realistic and violates the most basic
requirements for physical plausibility.

\section{Constant gravitational potential}

It has been shown that a necessary and sufficient condition for
isothermal behaviour, namely an inverse square fall off of the
density and pressure, is a constant spatial gravitational potential
$\lambda$. This is valid in Einstein theory and the more general
Lovelock theory \cite{NKD}. But what are the consequences of a
constant potential in $f(R,T)$ theory? We now examine this question.

Setting $Z = k$ for some constant $k$ in the isotropy equation
(\ref{5c}) gives \be e^{\nu} = c_2 r^{2-2 \sqrt{2-k}} \left(c_1+r^{2
\sqrt{2-k}}\right){}^2 \label{71}
 \ee for the remaining temporal
potential. Note that $k$ is now restricted through $0 < k < 2$.

Introducing the substitutions $a_1 = 2(\chi + 4\pi)$ and $a_2 = 8\pi
+ 3\chi$ the density and pressure are given by \beq \rho &=& \frac{2
\chi \left(\sqrt{2-k} \left(1-\frac{2 c_1}{c_1+r^{2
\sqrt{2-k}}}\right)+k\right)+8 \pi (k-1)}{8 k (\chi +2 \pi ) (\chi
+4 \pi )} \label{72a} \\ \n \\
p&=& \frac{2 a_2 \left(c_1(1- \sqrt{2-k})+\left(1+\sqrt{2-k}\right)
r^{2 \sqrt{2-k}}\right) - (k-1) a_1(c_1+r^{2 \sqrt{2-k}})}{8 k (\chi
+2 \pi ) (\chi +4 \pi )(c_1+r^{2 \sqrt{2-k}})} \n \\ \label{72b}
 \eeq
respectively while the sound speed has the remarkably simple
constant  value \be \frac{dp}{d\rho} = \frac{8 \pi }{\chi }+3
\label{73} \ee and to ensure causal behaviour $0 < \frac{dp}{d\rho}
< 1$ it is demanded that $\chi$ obeys $-4 \pi <\chi <-\frac{8\pi}{3}
.$

The expressions governing the energy conditions have the form \beq
\rho - p &=& \frac{c_1
\left(k+\sqrt{2-k}-2\right)+\left(k-\sqrt{2-k}-2\right) r^{2
\sqrt{2-k}}}{2 k (\chi +2 \pi ) \left(c_1+r^{2 \sqrt{2-k}}\right)}
\label{74a} \\ \n \\
\rho + p &=& \frac{\left(\sqrt{2-k}+1\right) r^{2 \sqrt{2-k}}-c_1
\left(\sqrt{2-k}-1\right)}{k (\chi +4 \pi ) \left(c_1+r^{2
\sqrt{2-k}}\right)} \label{74b} \\ \n \\
\rho + 3p &=& \frac{\chi  \left(\left(-k+5 \sqrt{2-k}+6\right) r^{2
\sqrt{2-k}}-c_1 \left(k+5 \sqrt{2-k}-6\right)\right)}{2 k (\chi +2
\pi ) (\chi +4 \pi ) \left(c_1+r^{2 \sqrt{2-k}}\right)} \n \\ \n
\\ &+& \frac{4 \pi \left(\left(-k+3 \sqrt{2-k}+4\right) r^{2
\sqrt{2-k}}-c_1 \left(k+3 \sqrt{2-k}-4\right)\right)}{2 k (\chi +2
\pi ) (\chi +4 \pi ) \left(c_1+r^{2 \sqrt{2-k}}\right)}\label{74c}
\eeq The active gravitational mass is calculated as  \be m =
\frac{r^3 \left(-2 \sqrt{2-k} \chi  \, _2F_1\left(1,\frac{3}{2
\sqrt{2-k}};1+\frac{3}{2 \sqrt{2-k}};-\frac{r^{2
\sqrt{2-k}}}{c_1}\right)+\left(k+\sqrt{2-k}\right) \chi +4 \pi
(k-1)\right)}{12 k (\chi +2 \pi ) (\chi +4 \pi )} \ee where $_2F_1$
is the familiar hypergeometric function.

In view of the complexity of the expressions for the dynamical
quantities we conduct a qualitative study with the aid of graphical
plots. The following parameter values have been used to generate the
plots $c_1 = 1$, $c_2 = 2$ and $k = 1.5$. Additionally we consider
three different values for the coupling $\chi$ namely $\chi = - 12$
(thick curve), $\chi = - 10$ (thin curve) and $\chi = 0$ (dashed
curve) - the last  is the Einstein case.

\begin{figure}
 \includegraphics[width=6cm]{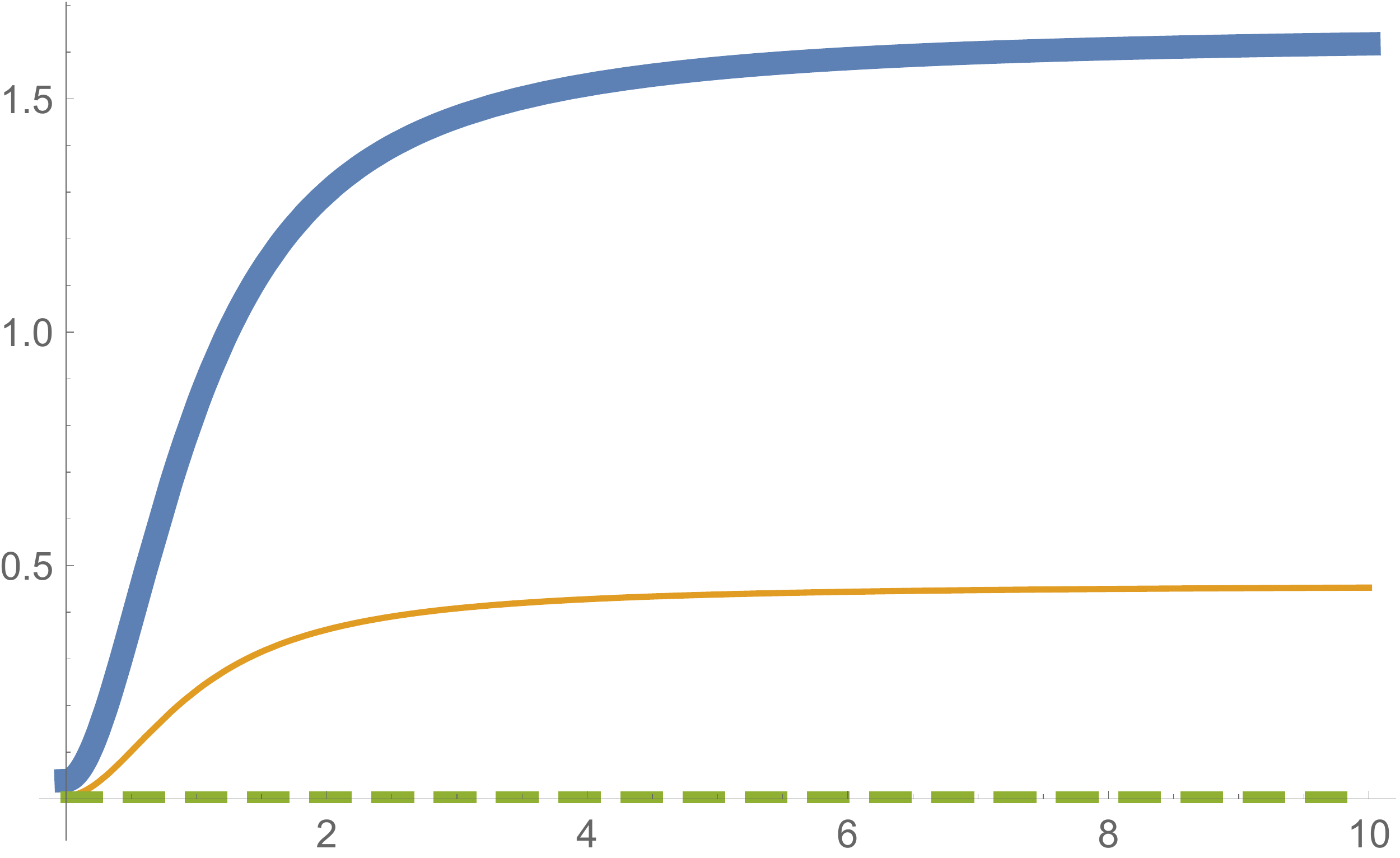}\\
  \caption{Density versus radial value $r$}\label{Fig 1}
\end{figure}

\begin{figure}
 \includegraphics[width=6cm]{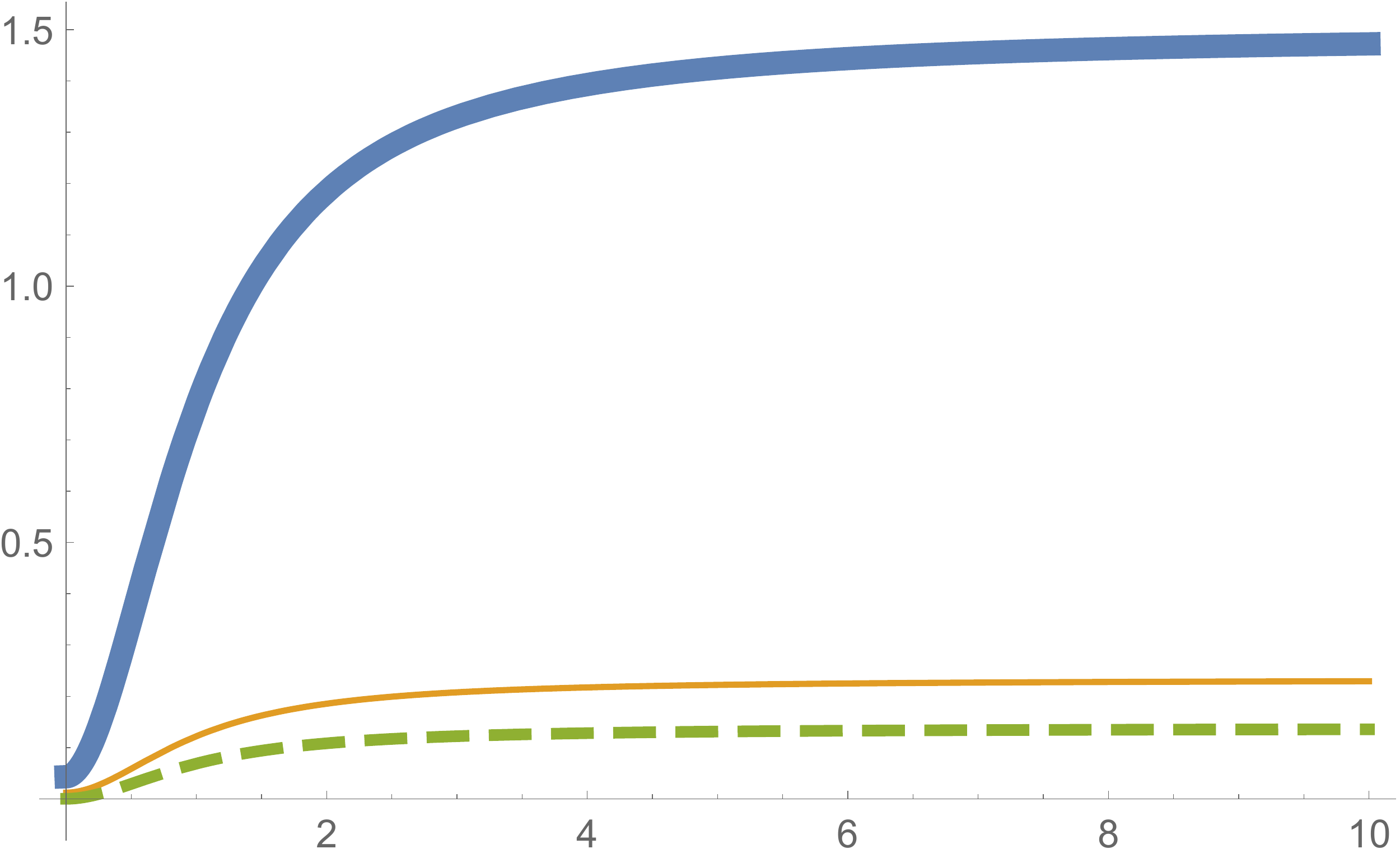}\\
  \caption{Pressure versus radial value $r$}\label{Fig 2}
\end{figure}

\begin{figure}
 \includegraphics[width=6cm]{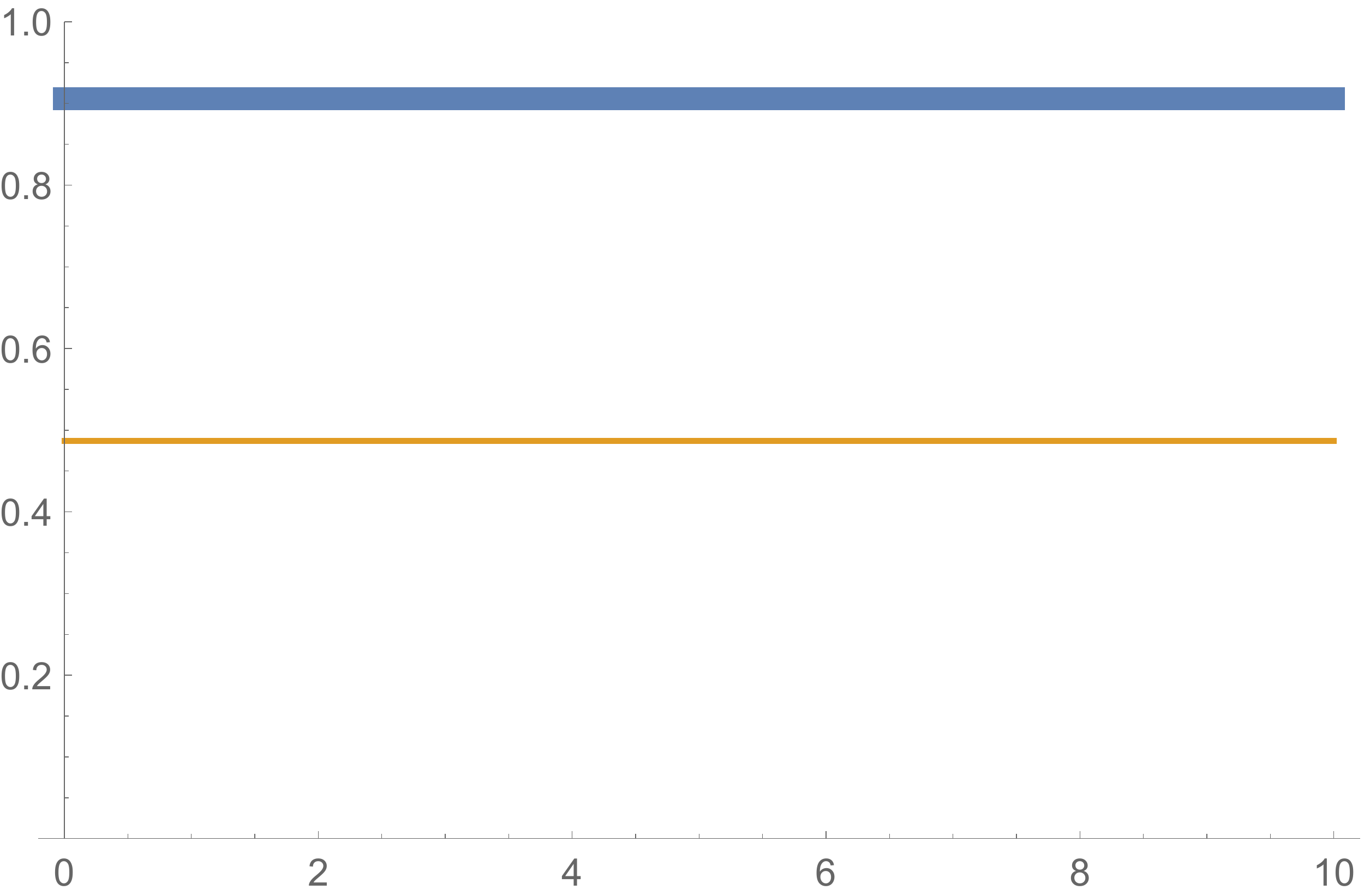}\\
  \caption{Sound speed squared versus radial value $r$}\label{Fig 3}
\end{figure}

\begin{figure}
  \includegraphics[width=6cm]{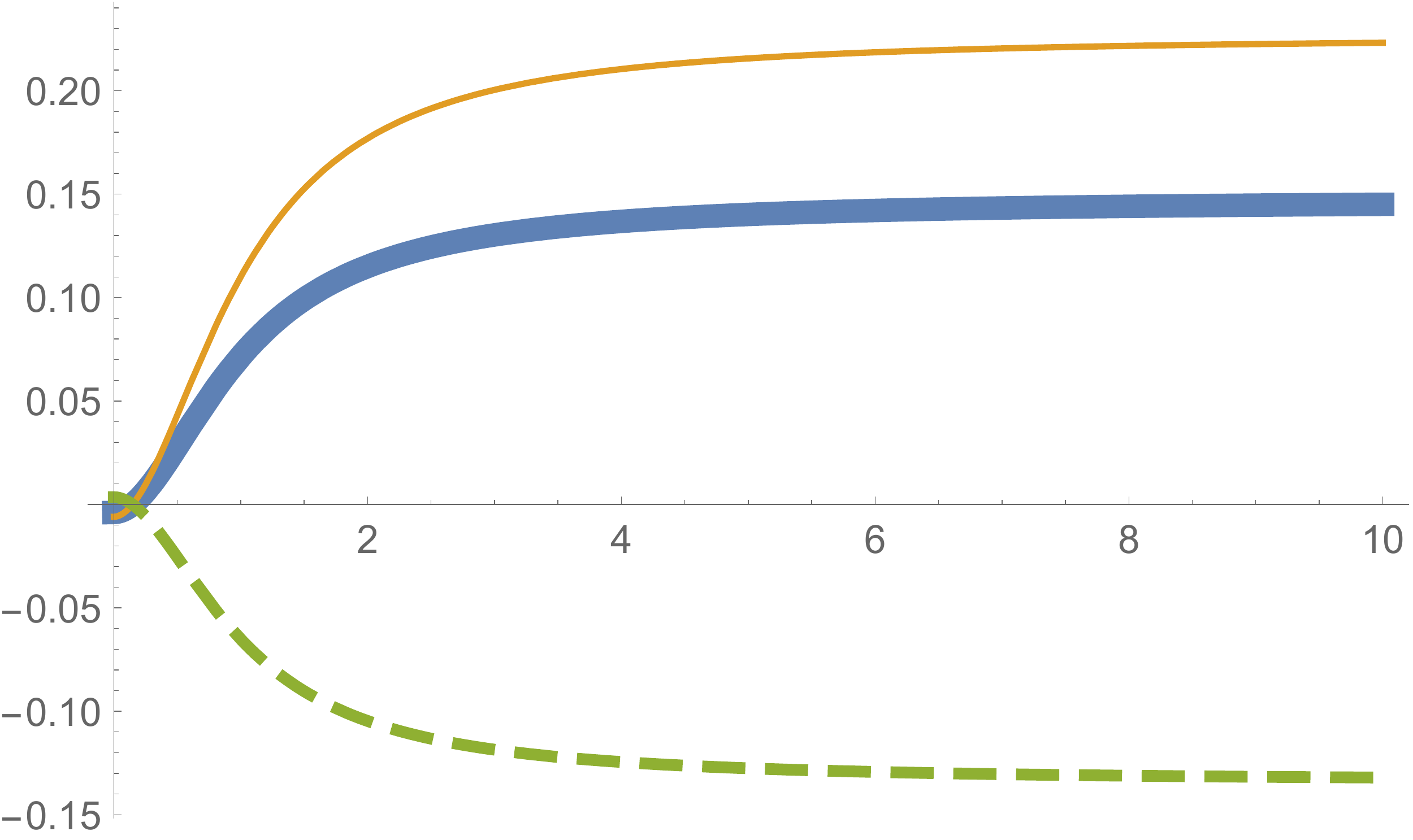}\\
  \caption{Weak energy condition versus radial value $r$}\label{Fig 4}
\end{figure}

\begin{figure}
  \includegraphics[width=6cm]{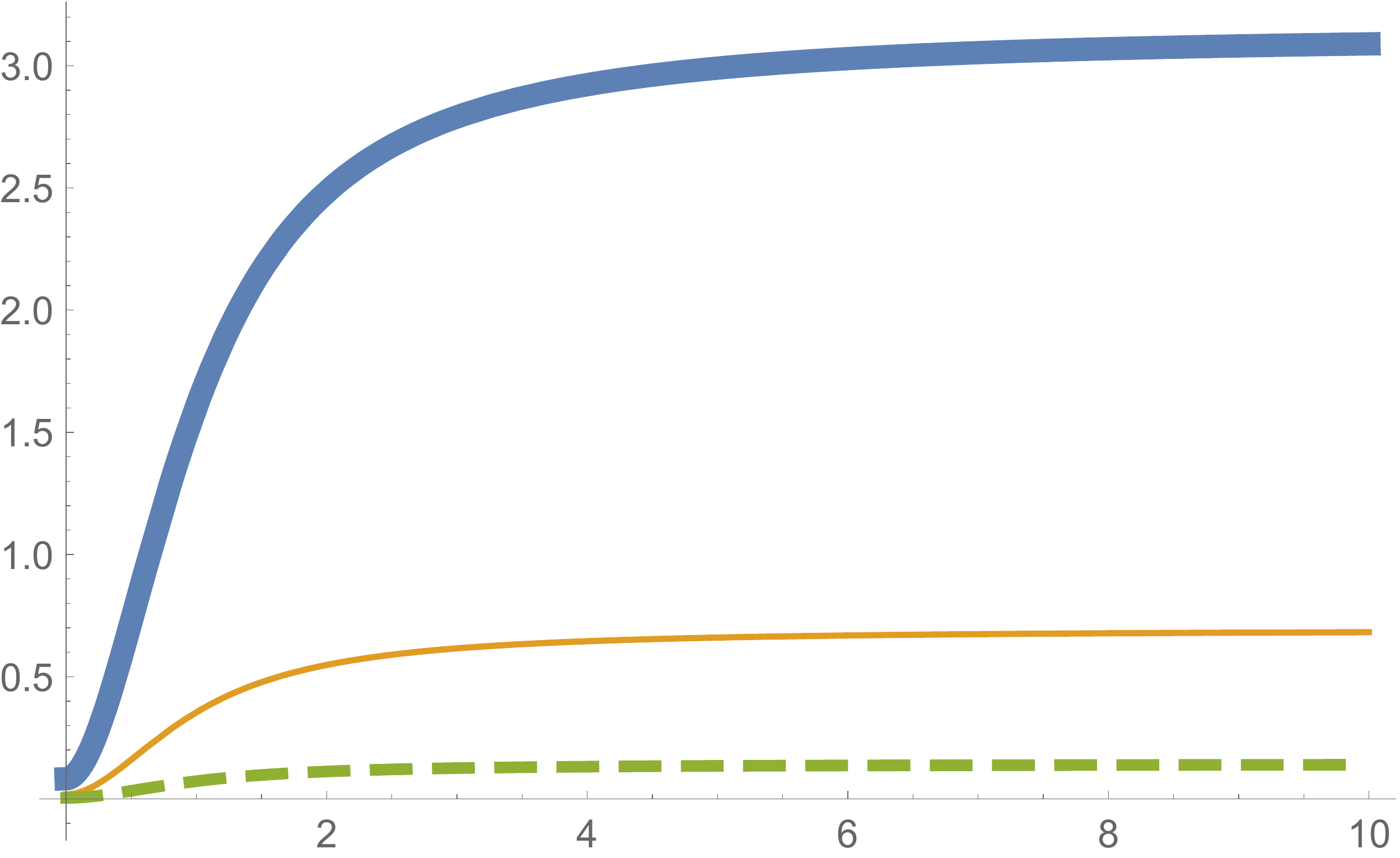}\\
  \caption{Strong energy versus radial value $r$}\label{Fig 5}
\end{figure}

\begin{figure}
  \includegraphics[width=6cm]{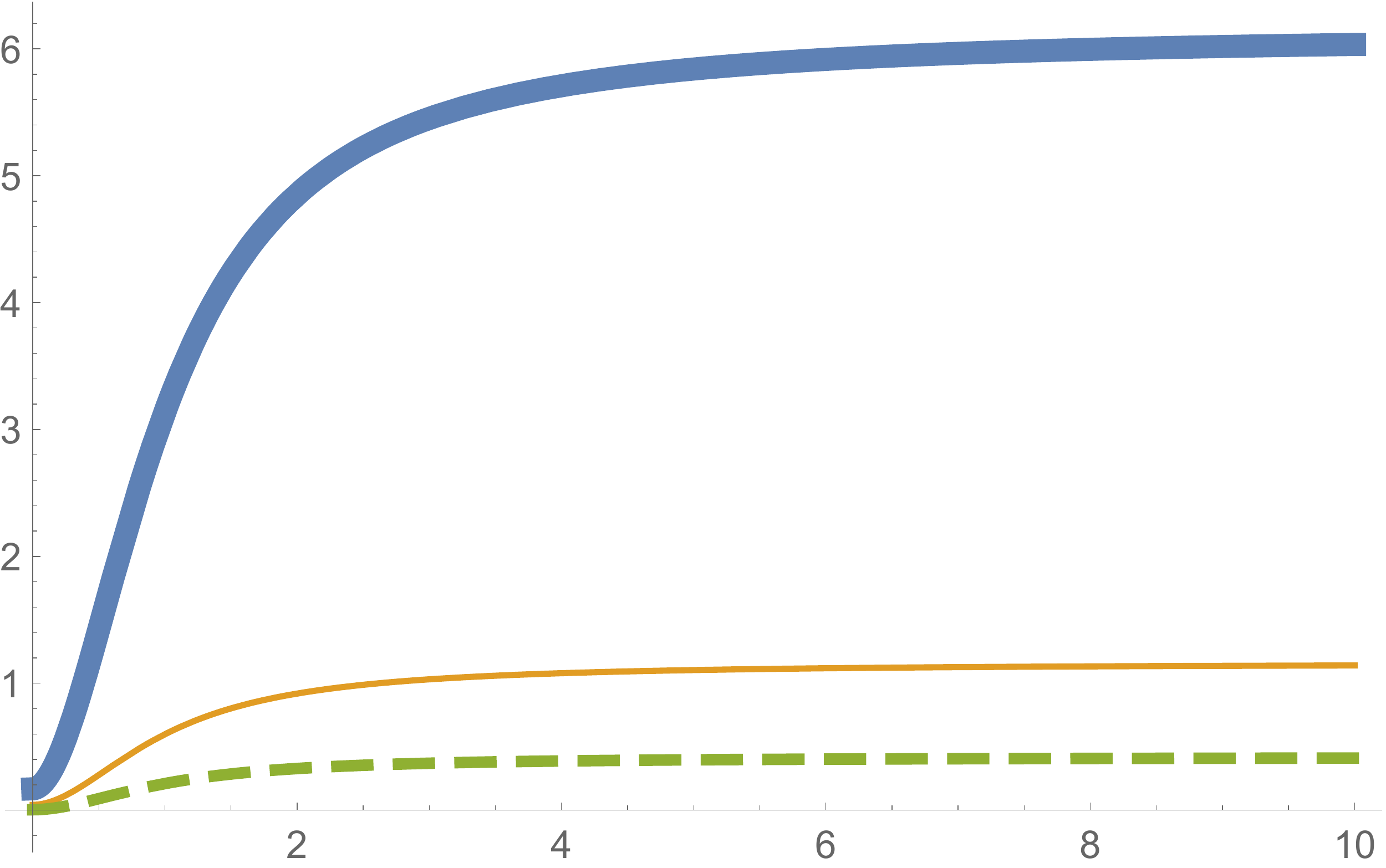}\\
  \caption{Dominant energy condition versus radial value $r$}\label{Fig 6}
\end{figure}

\begin{figure}
  \includegraphics[width=6cm]{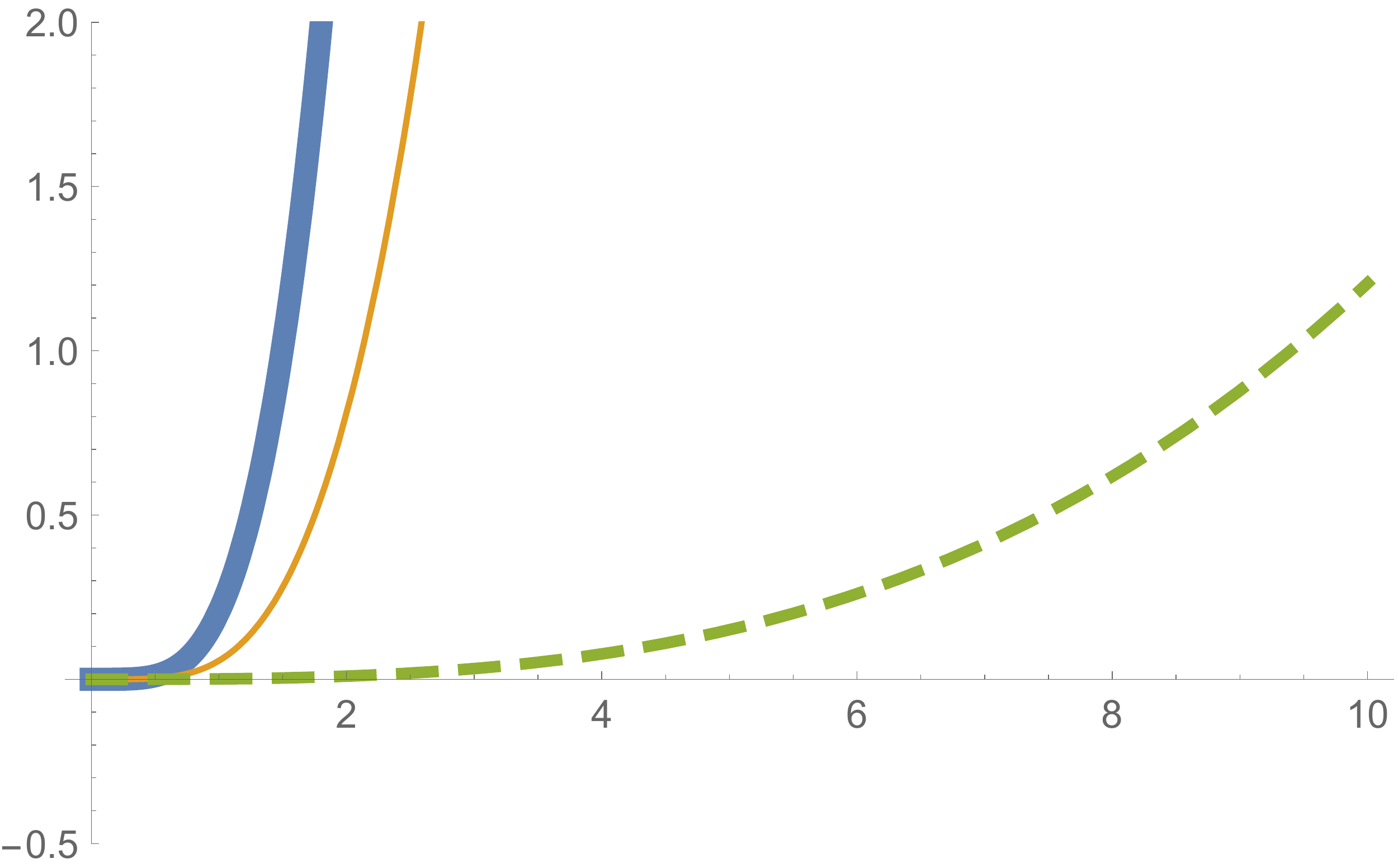}\\
  \caption{Mass versus radial value $r$}\label{Fig 7}
\end{figure}

Analysis of the plots: Figures 1  and 2 display the density and
pressure profile respectively and it can be observed that in all
cases of $\chi$ the curves are positive and increasing. The absence
of a surface of vanishing pressure is evident and is characteristic
of isothermal fluids. However, neither the density nor pressure
appear to obey the inverse square law fall off from the center.
Figure 3 depicts only curves for the $f(R,T)$ cases as the Einstein
case is meaningless. the sound speed values are in the acceptable
range of 0 to 1 to prevent a violation of causality. While the weak,
strong and dominant energy conditions (Fig 4, 5, 6) appear to be
well behaved for the $f(R,T)$ cases, the weak energy condition is
violated for the Einstein case. Finally the plot of the mass profile
(Fig 7) is reasonable. The mass increases more rapidly in the case
of the $f(R,T)$ theory than compared to its Einstein counterpart. In
summary, the $(f(R, T)$ case displays more pleasing physical
behavior than the Einstein case.

\section{Equation of state}

In the Einstein framework, imposing the equation of state $p =
\alpha \rho$ determines a relationship between the metric potentials
$\nu$ and $\lambda$. It is possible to isolate $\nu'$ and substitute
this into the equation of pressure isotropy - also an equation
connecting $\nu$ and $\lambda$. The caveat in this approach is that
the resulting nonlinear differential equation is difficult to
integrate and to date no unique general solution is known. An
alternative approach is to specify one of the four variables $\nu$,
$\lambda$, $p$ or $\rho$ and then to solve the system to reveal the
remaining three. Finally if the density or pressure equation is
solvable for $r$ in terms of $\rho$ or $p$ then a linear barotropic
equation can easily be determined albeit that the expressions are
lengthy. For example, if the density profile is prescribed in such a
way that the resulting  equation can be arranged as  a polynomial
equation up to quartic order in $r$, then the equation can be solved
for $r$ in terms of $\rho$. Substituting $r$  in the expression for
$p$ gives the equation of state. This equation of state is clearly
not the most general one for $p = \alpha \rho$ but represents a
special case. For example see the seminal work of Tolman
\cite{tolman} wherein some Tolman models do indeed display equations
of state.   Note that in the Einstein case, specifying the density
is tantamount to specifying the potential $\lambda$ as the $G^t_t =
T^t_t$ equation only contains $\lambda$ and $\rho$ and is well known
that the left-hand side may be expressed as an entire derivative.
This is not the case in the $f(R, T)$ scenario where both $\nu$ and
$\lambda$ appear in the same equation with $\rho$. For this reason
the incompressible fluid (constant density) solution is still
unknown in $f(R, T)$ gravity. However, there is some extra latitude
present through the constant $\chi$ and an equation of state may be
determined as shall be demonstrated below.

Imposing the equation of state $p = \alpha \rho$ results in the
relationship \be r \nu' = (\chi - \alpha a_2)r\lambda' - (\alpha
+1)a_1 (e^{\lambda} -1) \label{71}  \ee expressing $\nu$ in terms of
$\lambda$. Substituting (\ref{71}) into the isotropy equation
(\ref{5c}) generates the differential equation \beq &&(\alpha +1)^2
a_1^2 \beta^4+\beta^2 \left((\alpha +1) a_1 (\alpha a_1 +a_1
-4)+(\alpha +1) a_1 r (2 \alpha a_2 -2 \chi -1) \beta '-4\right)
 \n \\ \n \\ && +r
\beta  \left(2 r (\chi -\alpha a_2) \beta ''-\beta ' ((\alpha +1)
a_1 (2 \alpha a_2-2 \chi +1)-2 \alpha a_2 +2 \chi +2)\right)  \n
\\ \n \\ && -2 ((\alpha +1) a_1 (\alpha a_1+a_1 -2)-2) \beta
^3+r^2 (\alpha a_2 -\chi ) (\alpha a_2 -\chi +3) \beta '^2 = 0
\label{81}
\eeq
governing the behaviour of $e^{\lambda} = \beta$. Obtaining the general solution to (\ref{81}) has proved elusive in view of the nonlinearity. The method of Lie group analysis was invoked however no symmetries could be detected immediately. However, on careful observation it is seen that in some cases (\ref{81}) may be solved explicitly.

For the special case $\alpha = \frac{\chi}{a_2}$ the isotropy
equation becomes \be \beta (\beta -1)  \left(c_1^2 \beta +
c_2\right) -r\beta' \left(c_1 (\beta +1) +2  \right) = 0 \label{82}
\ee where we have redefined  $c_1 = (\alpha + 1)a_1$ and  $c_2 =
-(\alpha +1) a_1 (\alpha a_1 + a_1-4)+4$. Dividing throughout by the
first term on the left we may rearrange  equation (\ref{82}) to the
form \be \beta' \left( -\frac{(c_1 + 2)}{c_2 \beta} + \frac{2(c_1 +
1)}{(c_1^2 + c_2)(\beta - 1)} + \frac{c_1^3(2c_1 +  c_1^2 - c_2
)}{c_2(c_1^2 + c_2)(c_1^2 \beta + c_2)} \right) = \frac{1}{r}
\label{83} \ee with the help of partial fractions. The solution by
quadratures may now be obtained implicitly as \be \frac{(\beta
-1)^{\frac{2(c_1 + 1)}{(c_1^2 + c_2)}} (c_1^2 \beta +
c_2)^{\frac{c_1( c_1^2 +2c_1   - c_2)}{c_2 (c_1^2 + c_2)}}
}{\beta^{\frac{c_1 + 2}{c_2}}} = Kr \label{84} \ee where $K$ is a
constant of integration. Equation (\ref{84}) is essentially an
algebraic equation in $\beta(r)$. Clearly for judicious choices of
the constants $c_1$ and $c_2$, equation (\ref{84}) may be solved
explicitly to find the gravitational potential function $\beta$.

As an example, consider the choice $c_1 = -2$ and consequently $c_2
= - 8$ follows. Now from $\frac{c_1}{\alpha + 1} = 2(\chi + 4\pi)$
and the original assumption $\alpha = \frac{\chi}{a_2}$ we solve
simultaneously and obtain the pair \be \{(\chi, \alpha)\} =
\left\{\left(\frac{ \left(\mp \sqrt{9+16 \pi +64 \pi ^2}-24 \pi
-3\right)}{8}\right)  , \left(\frac{\pm\sqrt{9+16 \pi +64 \pi
^2}-3}{8 \pi }\right)\right\} \ee or given approximately numerically
as $\{(\chi, \alpha)\} = \{ (-13.0854, 0.926495),( -6.51411,
-1.16523)\} $. We must discard the negative value of $\alpha$ since
the causality criterion $0 < \alpha < 1$ will be violated. However,
note that we are able to obtain the value $\alpha = 0.926$ which
indeed guarantees a subluminal sound speed. For this choice of $c_2$
equation (\ref{84}) is solvable and the metric potential evaluates
to \be \beta = e^{\lambda} = \frac{1-4K^2r^2}{1-2K^2r^2}  \ee which
corresponds to the Vaidya-Tikekar \cite{vt} spheroidal geometry
utilised to model superdense relativistic stars. In order to
determine the remaining gravitational potential it is prudent to
introduce the transformations $x = 2K^2r^2$, $Z(x) = e^{-\lambda}$
and $e^{\nu} = y^2 (x)$ whence the equation of pressure isotropy
assumes the form \be 4x^2Z\ddot{y} + 2x^2\dot{Z}\dot{y} + (\dot{Z}x
- Z +1)y=0 \label{87} \ee and is now a second order linear
differential equation in $y$. Inserting $Z=\frac{1-x}{1-2x}$ into
(\ref{87}) generates the potential \be y = c_1 \sqrt{1-x}+2 c_2
\left(\sqrt{1-2 x}- \sqrt{2(1-x)} \log \left(2 \sqrt{1-x}+\sqrt{2-4
x}\right)\right) \ee or in the canonical form \be e^{\nu} = c_1 v_1
+2 c_2 \left(v_2- \sqrt{2}v_1 \log \left(2
v_1+\sqrt{2}v_2\right)\right) \ee where we have put $v_1 =
\sqrt{1-2K^2r^2}$ and $v_2 = \sqrt{1-4K^2r^2}$. The pressure and
density are given by \beq
p &=& \alpha \rho  \n \\ \n \\
&=& \frac{2\alpha K^2r^2}{8 K_1 K_2 v_2^2} \left(  \frac{ \chi v_1
\left(2\sqrt{2} c_2 v_2 v_3 \log v_3 - c_1\right)}{\sqrt{2}\left(c_1
v_1+2 c_2 \left(v_2-\sqrt{2}v_1 \log v_3\right)\right)}-\frac{2 (3
\chi +8 \pi )}{v_2^2}- 2 K_1  \right) \n \\  \label{88} \eeq where
we have made the further simplifications $v_3 = \left(2
v_1+\sqrt{2}v_2\right)$, $K_1 = \chi + 4\pi$ and $K_2 = \chi +
2\pi$. Now we have a complete model with Vaidya--Tikekar \cite{vt}
geometry and linear barotropic equation of state $p = \alpha \rho$.
a defect in this model is that there exists an essential singularity
at $r= \pm \frac{1}{2K}$. While the presence of the singularity is
undesirable, it may not be a generic feature of this model. Suitable
constants $c_1$ and$c_2$ may yet exist that support a well behaved
cosmological model. Interestingly, the vanishing of the pressure for
a finite $r$ is possible allowing for the interpretation of this
model as a bounded astrophysical distribution.

\section{Relaxing the equation of state}

Finally we consider the case where the density displays an inverse
square-law fall-off but we refrain from imposing an equation of
state. That is the system of field equations is now completely
determined and the resulting solution should be inspected for an
equation of state. It turns out that equation (\ref{5a}) allows us
to write $\nu'$ in terms of $\lambda$ and its derivative. When this
form is substituted into the isotropy equation (\ref{5c}) the
resulting differential equation proves intractable to solve. Note
that this situation does not arise in the standard Einstein gravity
since on setting $\chi = 0$ for the Einstein case, (\ref{5a}) can be
solved explicitly for $\lambda$ in terms of $r$. This has been amply
demonstrated by Dadhich {\it{et al}} \cite{dad-hans, NKD} for the
Einstein case and its generalization pure Lovelock theory.

\section{Conclusion}

We have analysed the isothermal property in the framework of
$f(R,T)$ theory. Demanding an inverse square fall-off of the density
and the equation of state $p = \alpha \rho$ yielded an exact model
where the proportionality constant $\alpha$ is expressed in terms of
the coupling constant $\chi$. For stability and to prevent
super-luminal behavior of the fluid the value of $\chi$ was
constrained to a certain negative window. On setting $\chi = 0$ we
regain the Saslaw {\it{et al}} model for standard Einstein gravity
and we discover that it is not physically reasonable. In contrast,
the $f(R, T)$ model displayed the necessary features corresponding
to expectations, namely a positive definite density and pressure and
a sound speed obeying causality. While it is known that a constant
spatial potential guarantees isothermal behaviour in the Einstein
case and its generalization Lovelock gravity, such a prescription
behaves completely differently in the $(f(R, T)$ gravity framework.
Dropping the inverse square law requirement and requiring an
equation of state, the $f(R, T)$ model is indeed solvable in at
least one special case. We have given a prescription to determine
other models which satisfy the field equations and the equation of
state. The case of an inverse square fall-off of the density without
an equation of state did not yield an exact solution.



\bibliography{basename of .bib file}

\end{document}